\documentclass{optica-article}

\journal{opticajournal} 

\articletype{Research Article}

\usepackage{float}
\usepackage{graphicx}
\usepackage{svg}
\usepackage{tikz}
\usetikzlibrary{shapes.geometric, arrows}
\graphicspath{{images/}}
\usepackage{lineno}

\usepackage{booktabs}

\tikzstyle{startstop} = [rectangle, rounded corners, minimum width=3cm, minimum height=1cm,text centered, draw=black, fill=white]
\tikzstyle{io} = [trapezium, trapezium left angle=70, trapezium right angle=110, minimum width=3cm, minimum height=1cm, text centered, draw=black, fill=white]
\tikzstyle{process} = [rectangle, minimum width=3cm, minimum height=1cm, text centered, draw=black, fill=white]
\tikzstyle{decision} = [diamond, minimum width=3cm, minimum height=1cm, text centered, draw=black, fill=white]
\tikzstyle{arrow} = [thick,->,>=stealth]

\begin{document}
\title{Ultra-broadband room-temperature Fourier transform spectrometer with watt-level \\ power consumption}
\date{\vspace{-5ex}}

\author{Jakub Mnich,\authormark{1,*} 
Johannes Kunsch,\authormark{2} Matthias Budden,\authormark{3} \\ Thomas Gebert,\authormark{3} Marco Schossig,\authormark{4} Jarosław Sotor,\authormark{1}\\ and Łukasz A. Sterczewski\authormark{1,5}}

\address{\authormark{1}Laser and Fiber Electronics Group, Faculty of Electronics, Photonics and Microsystems, Wrocław University of Science and Technology, Wybrzeże Stanisława Wyspiańskiego 27, 50-370 Wrocław, Poland\\
\authormark{2}LASER COMPONENTS Germany GmbH, Werner-von-Siemens-Str. 15, 82140 Olching, Germany\\
\authormark{3}WiredSense GmbH, Luruper Hauptstr. 1, 22547 Hamburg, Germany\\
\authormark{4}Infrasolid GmbH, Gostritzer Straße 61-67, 01217 Dresden, Germany}

\email{\authormark{*}jakub.mnich@pwr.edu.pl} 
\email{\authormark{5}lukasz.sterczewski@pwr.edu.pl}

\begin{abstract*}
    Fourier-transform infrared spectroscopy (FTIR) has matured into a versatile technique with relevance for environmental monitoring, pharmaceutical research, and food safety applications. However, compared to other spectroscopic methods, it experiences slower progress in terms of power optimization, miniaturization, and adoption by industry. To overcome this limitation, we developed an ultra-broadband room-temperature FTIR instrument relying on commercially available components that offers a spectral coverage from 1.6~$\upmu$m to 31~$\upmu$m (9.7--190~THz) without changing optics at a single-Watt-level of electrical power consumption. To demonstrate the capabilities of the instrument, we measured atmospheric species in multiple spectral regions with better than 1.5~cm\textsuperscript{-1} resolution.
\end{abstract*}

\section{Introduction}
The Michelson interferometer lying at the heart of most FTIR spectrometers is a XIX-th century invention that has played a pivotal role in many scientific discoveries. Arguably, the two most famous testaments to this are the disproval of the existence of luminiferous ether in the Michelson-Morley experiment~\cite{michelson1887relative}, and the observation of gravitational waves at the Laser Interferometer Gravitational-Wave Observatory (LIGO)~\cite{abbott2016gw151226}. For spectroscopic purposes, Michelson's invention has required more than half a century to reach commercial maturity. The first widely-available FTIR instrument with a 2~cm\textsuperscript{-1} (60~GHz) resolution, sufficient for most solid-state and liquid analytes, was released by Perkin Elmer (KBr~Model~137) in March of 1958~\cite{kbr_model_137, periknelmer_60years}. Since then, such instruments have become workhorses in many areas of modern science: chemistry, biology, pharmacy, optics, physics, geology, climate etc.~\cite{griffiths}. Despite the well-understood benefits that would come from the widespread adoption of FTIR in industry, many practical obstacles have impeded this process including high cost, cooling requirements for low-noise photodetection, limited maintenance-free operation of the source, and susceptibility to interferometer misalignment due to environmental factors. Although miniaturized FTIR instruments already exist, their severely limited performance confines them to critical applications in threat detection~\cite{food_FTIR, FTIR_coffee_spices, advances_portable_ftir, crocombe_portable_2021, crocombe_portable_2021_2}. In general, FTIRs remain mostly limited to use in specialized laboratories. Over the last decades, substantial effort has been poured into providing complementary technologies for FTIR spectroscopy including new detectors, sources, beam splitters and optomechanical devices~\cite{portable_spectroscopy} but fundamental limitations on FTIRs capabilities persist~\cite{ftir_market_strategy}. The result is that none of the current FTIR technology platforms can be considered truly universal and instruments compete with each other based on secondary characteristics: sample interfaces, analytic capabilities of the software, physical form factor, etc..

To provide a viable solution for FTIR spectroscopy on a larger scale, in this work, we propose to employ a commercially available thin-membrane pyroelectric detector in tandem with a~ceramic powder-coated (CPC) thermal source to achieve a broadband spectral coverage from 1.6~to~31~\textmu m with 23~dB of maximum dynamic range over seconds, while offering excellent secondary characteristics. These include low-power consumption (the source consumes only 1.3~W of electrical power) full room-temperature operation without any cooling of the detector and source, fast scan speeds (demonstrated 2~mm/s, limited by the opto-mechanical stability of the translation stage), and potential for further miniaturization supported by a rugged nature of optical components including an integrated beam splitter with a germanium layer encapsulated between two identical KBr plates that removes the need for a separate dispersion compensation element. Instead of a conventional HeNe laser, a telecommunication-wavelength distributed feedback (DFB) laser diode is employed as a reference source to track the instantaneous optical path difference via a Hilbert transform beyond the conventional fringe counting limit. The digital, oversampling routine of correcting translation stage positioning errors enables one to use constant-rate data acquisition systems incapable of synchronizing the interferogram sampling process to an external trigger signal (see Section \ref{DataAcqAndProcSection}). 

Key enabling technologies for this work are identified, discussed, and compared with existing solutions. We evaluate the performance of the instrument in two steps: first by measuring native source spectra to assess the spectral coverage and dynamic range followed by practical absorption measurements of H\textsubscript{2}O and CO\textsubscript{2} in multiple spectral regions. We confirm the validity of our spectroscopic results using the HITRAN2020 database~\cite{hitran}.

\section{Materials and methods}
\subsection{Setup}
\begin{figure}[b!]
	\centering \includegraphics[width=11cm]{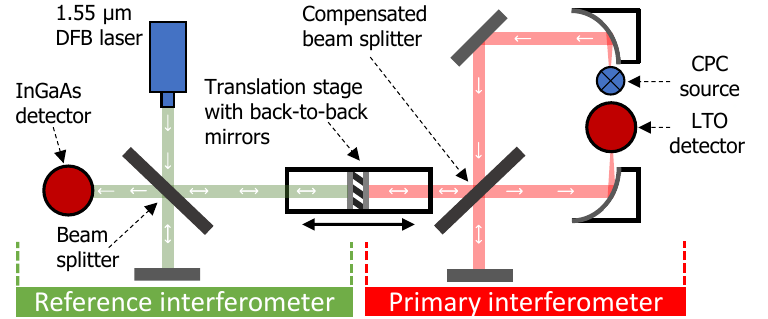} 
	\caption{Experimental setup of the FTIR spectrometer; DFB~-~distributed feedback, LTO~-~LiTaO\textsubscript{3}, CPC~-~ceramic powder-coated;}
	\label{fig:setupDiagram}
\end{figure}
\noindent To facilitate the characterization of different experimental configurations, a dedicated FTIR setup was assembled on an optical breadboard. As shown in Fig.~\ref{fig:setupDiagram}, the setup includes two Michelson interferometers: the primary -- responsible for producing interferograms (IGMs) of broadband radiation sources, and the reference -- responsible for producing a reference sinewave-like positioning signal from a narrow-band laser source (see Section \ref{DataAcqAndProcSection}). Typically, both beams share the same interferometer, but since this setup is designed to allow for easy replacement of the beam splitter, separating them makes realignment easier. It also allows for devoting more aperture of the primary beam splitter to the measured beam, thus increasing the amount of light reaching the detector. The compensated beam splitter (Omega Optical LLC, USA) utilizes a~germanium-based surface sandwiched between two potassium bromide (KBr) windows with a~humidity-protective anti-reflective coating. This substrate material offers an extremely broadband transmission window with a long-wavelength limit of $\sim$30~$\upmu$m.

The primary beam is collimated with an off-axis parabolic (OAP) mirror and then either fed directly into the optical system or first passed through a~gas cell containing the analyte. Another OAP is then used at the exit of the interferometer to focus the beam from the diameter of about 1"~down to \textless~2~mm to best utilize the active area of the detector (2.0~$\times$~2.0~mm$^2$).    

\subsection{Detector}
A typical material used for high-performance pyroelectric detectors in FTIRs is deuterated \textsc{l}-alanine doped triglycene sulphate~(DLATGS), which is a derivative of triglycene sulphate~(TGS) with additions into its lattice serving two purposes: increasing its thermal stability and $D^{*}$ of the detector. Despite these improvements, the Curie temperature of DLATGS is still around~60\textdegree C and thus such detectors are most often paired with thermoelectric coolers. Moreover, DLATGS exhibits a significant thermal non-linearity of 2.5~\%/\textdegree C which was shown to cause up to a 1\% voltage error with irradiances as low as 7~mW/mm\textsuperscript{2}~\cite{linearity_DLATGS}. Such properties are highly undesirable in instruments designed for precise chemometric applications. The DLATGS material is also hygroscopic and will degrade upon exposure to ambient air; hence, it requires a hermetic package with a window, which in turn leads to a limited spectral bandwidth of detectors otherwise fundamentally capable of full coverage from the UV, through VIS, NIR, and MIR up to the THz range. 

Detectors used in this work employ crystals of lithium tantalate (LiTaO\textsubscript{3} or LTO), which additionally finds application in other non-photonic devices like piezoelectric filters \cite{pyroelectric_review_2023}. Although LTO-based photodetectors have been known to be suitable for broadband thermal light spectroscopy, we would like to make a distinction between two types of device architectures: 
\begin{itemize}
	\item Typical bulk-crystal devices with an active area thicker than 20~\textmu m, whose fabrication relies on mechanical polishing \cite{stokowski_ion-beam_1976} of LTO. 
	\item Thin-membrane devices with an active area thinner than 20~\textmu m, which is usually fabricated via ion milling. This is because LTO cannot be grown as thin layers directly on a substrate due to bulk-grown crystals showing 3~to~20$\times$~better pyroelectric properties~\cite{schossig_dielectric_2014, magnetron_lto}.
\end{itemize}

This distinction is important as most noise sources in pyroelectric detectors are suppressed in thinner crystals, as first shown in 1976 by Stokowski et al.~\cite{stokowski_ion-beam_1976} in thinner-than-hair sensors. Therefore, thinning the sensor yields an improved responsivity ($D^{*}$). Thin-membrane LTO detectors are expected to show $D^{*}$ over 1$\times$10\textsuperscript{8}~cm~Hz\textsuperscript{1/2}~W\textsuperscript{-1}~\cite{pyroelectric_review_2023, stokowski_ion-beam_1976}. Another advantage is their compatibility with higher modulation frequencies due to a lower thermal mass. To facilitate handling such structures, in most cases only the sensing area is thinned by ion-milling while the surrounding crystal is left with a~few dozen~\textmu m of material to improve structural stability.

In our work, we evaluate the spectroscopic performance of the FTIR instrument using two thin-membrane (5~$\upmu$m thickness) LTO pyroelectric detectors differing in the characteristics of the integrated electronics (see Table~\ref{tab:detectors}). Pyroelectric detector (PR)~no.~1 is an off-the-shelf commercial device (Laser Components, Germany) designed to detect low-power signals; hence it is easily saturated by the evaluated IR source, while PR~no.~2 is an early-stage prototype with a responsivity ($\mathcal{R}$) better matched to the amplitude of IGMs produced by the setup. Both detectors exhibit atypically high \textit{f}\textsubscript{3dB} frequencies~\cite{pyroelectric_review_2023} and maintain $D^{*}$ close to or above 1$\times$10\textsuperscript{8}~cm~Hz\textsuperscript{1/2}~W\textsuperscript{-1} for $f$~\textgreater~1~kHz, with PR~no.~2 holding the current record for \textit{f}\textsubscript{3dB} among LTO-based detectors. Further improvements to \textit{f}\textsubscript{3dB} are possible at the expense of lower $D^{*}$.

\begin{table}[H]
\centering
	\begin{tabular}{l c c c c}
		& \textit{f}\textsubscript{3dB} [kHz]& $\mathcal{R}$\textsubscript{1kHz} [kV/W]& NEP\textsubscript{1kHz} [nW/Hz\textsuperscript{1/2}] & $D^{*}$\textsubscript{1~kHz}  [cm~Hz\textsuperscript{1/2}~W\textsuperscript{-1}]\\
        \hline
		PR no. 1 & 8  & 70    & 1.3 & 1.5$\times$10\textsuperscript{8}      \\
		PR no. 2 & 50 & 5     & 2.3 & 8.6$\times$10\textsuperscript{7}    \\
        \hline
	\end{tabular}
	\caption{Parameters of the evaluated LTO detectors: 3~dB modulation bandwidth (\textit{f}\textsubscript{3dB}), responsivity ($\mathcal{R}$) and noise equivalent power (NEP).}
	\label{tab:detectors}
\end{table}

LTO has a high Curie temperature in excess of~600\textdegree C and is highly resistant to chemical degradation. As a result, detectors made with it are capable of operating in full exposure to ambient conditions, which removes the need for protective windows and hermetic packaging. The thermal linearity of LTO detectors is more than an order of magnitude better than in the case of DLATGS with a temperature coefficient of~0.2~\%/\textdegree C~\cite{linearity_LTO}. Without any protective window, the presented LTO detectors should cover the full spectral range from UV up to THz, although both devices discussed in this paper are equipped with potassium bromide (KBr) windows for protection against acoustic noise and air fluctuations, as well as preventing damage during handling and storage.

\subsection{Source}
Despite the significant progress in the field of non-thermal infrared radiation sources, like quantum cascade lasers (QCLs)~\cite{cargioli2024quantum}, interband cascade lasers (ICLs)~\cite{meyer2020interband} and other, more exotic emitters like infrared supercontinuum~\cite{zorin2022advances}, black body thermal emission remains the dominant scheme for generating broadband radiation in FTIR spectrometers~\cite{next_gen_sources}. The performance of such elements is limited by the Planck radiation law, practical consequences of which can effectively be summarized in the following statements:
\begin{itemize}
	\item The only way to increase the total energy radiated from a given area of an ideal black body is to heat it up since it is proportional to the fourth power of the surface temperature as stated by Stefan–Boltzmann law.
	\item Heating a black body shifts its peak emission towards shorter wavelengths in accordance with Wien's displacement law. 
	\item A non-ideal black body cannot exceed an ideal black body's emission power at the same temperature which means its emissivity $\epsilon$ always satisfies $\epsilon(\lambda, T) < 1$.
\end{itemize}

The globar is arguably the most popular thermal source in FTIR spectroscopy, which typically constitutes a silicon carbide (SiC) heater operated between 1000\textdegree C and 1600\textdegree C, placing its peak emission wavelength between 1.5~\textmu m and 2.2~\textmu m. This is optimal for spectroscopy in the near-infrared (NIR) but results in less radiated power in the mid-infrared (MIR) and long-wavelength infrared (LWIR) range. The power consumption of a typical globar ranges from 20~to~40~W, similar to a light bulb, but there are no fundamental limitations in that regard. Operation at elevated temperatures can be assisted by active cooling schemes to prevent damage to components.

Lowering the temperature from 1600\textdegree C to 1000\textdegree C, aside from increasing the peak emission wavelength, lowers the peak spectral radiance $\sim$5~times. This is the price one pays for shifting the thermal sources' peak emission deeper into MIR. 

\begin{figure}[!t]
	\centering \includegraphics[width=13.5cm]{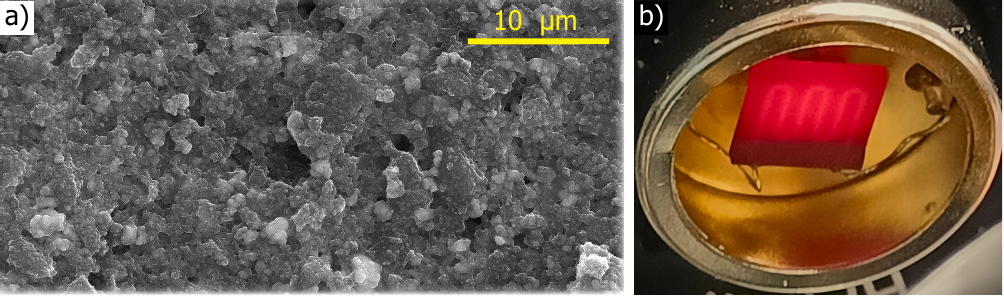} 
	\caption{a) SEM image of the ceramic coating of the emitter, showing the complex structure formed by the ceramic micro-powder. b) Close-up photo of the emitter's surface heated up to $\sim$650\textdegree C.} 
	\label{fig:nonoStruct}
\end{figure}

Instead of a high-power globar, we evaluated a prototype CPC thermal source (Infrasolid, Germany, see Fig.~\ref{fig:nonoStruct}b)~\cite{baliga_2024}, with a special ceramic particle black coating applied to the active surface (see Fig.~\ref{fig:nonoStruct}a). This source is designed to operate at surface temperatures ranging from 100~to~1000\textdegree C, which corresponds to peak emission between 2.3~\textmu m and 7.8~\textmu m. Lowered temperatures result in an extended source lifetime compared to globars and pose a lower fire risk. 

\begin{figure}[H]
	\centering 
    \includegraphics[width=0.9\textwidth]{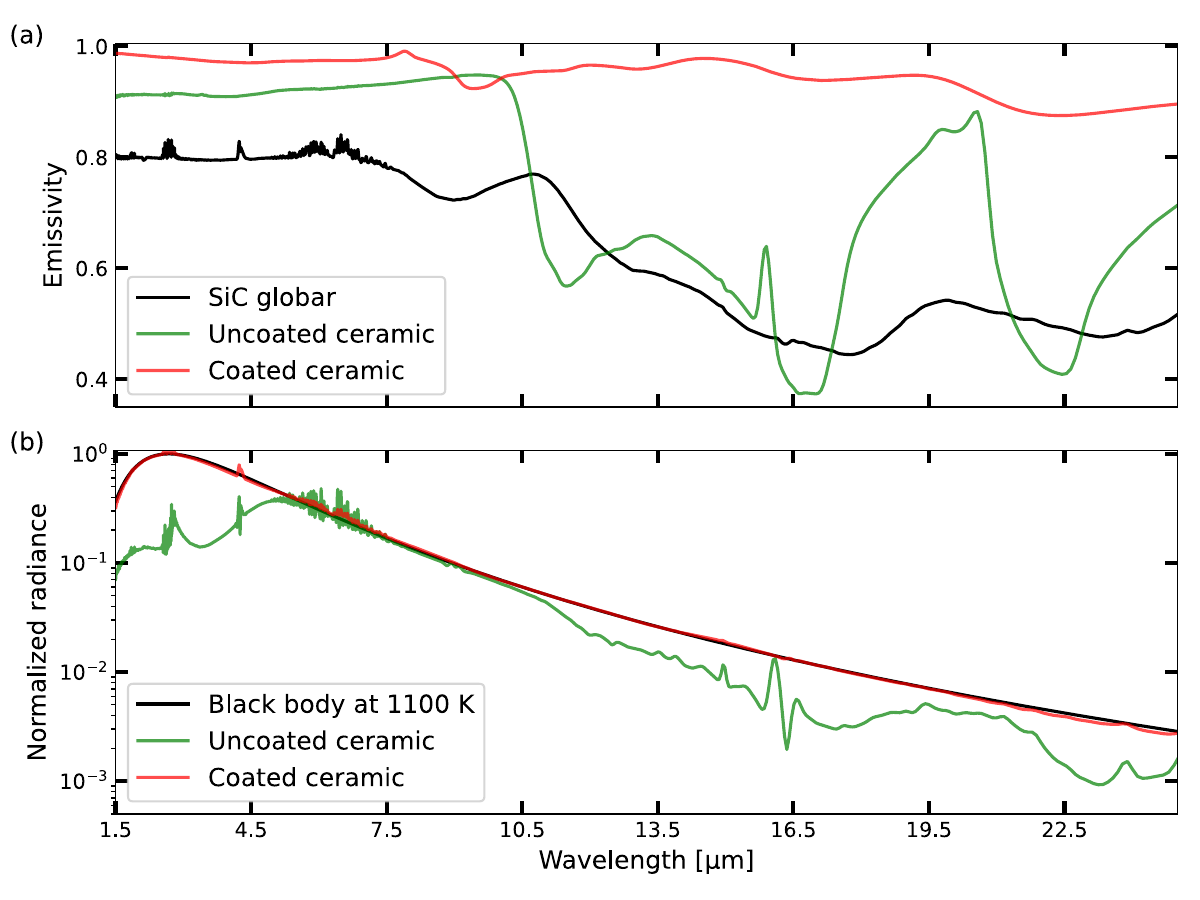}
	\caption{(a)~Comparison of room-temperature emissivity of a commercial silicon carbide (SiC) globar with an Al\textsubscript{2}O\textsubscript{3} ceramic substrate with and without a CPC coating; (b)~comparison of radiance from an ideal black body against coated and uncoated ceramic substrates.}
	\label{fig:infrasolid_radiance}
\end{figure}

The effect that the ceramic coating shown in Fig.~\ref{fig:nonoStruct}a has on the spectral radiance characteristics of an element is presented in Fig.~\ref{fig:infrasolid_radiance}a where its emissivity is compared against a commercial SiC globar and on Fig.~\ref{fig:infrasolid_radiance}b where the radiance of coated and uncoated ceramic substrate is compared against an ideal black body. The spectral radiance has been normalized to the peak of the ideal black body radiation curve. CPC source exhibits flatter broadband charactersitics and higher emissivity than the evaluated globar. Measurements were conducted by coupling the radiation into a Bruker Vertex 70v FTIR spectrometer with sources at the temperature of 1300~K. SiC globar consumed around 20~W of electrical power and the CPC source around 10~W \cite{guo_measurement_2019}. 

The radiance closely resembling that of an ideal black body offers multiple advantages: more broadband optical power emission, with relation to the peak radiance point, better efficiency in converting electrical into optical power, and the ability to operate at lower temperatures without sacrificing performance. These features allow one to reach deeper into MIR and THz wavelengths with good spectral dynamic range even accompanied by lower power consumption. Such emitters will be referred to as nanostructured thermal sources (NTS).

\subsection{Data acquisition and processing} \label{DataAcqAndProcSection}
\begin{figure}[H]
	\centering \includegraphics[width=0.75\textwidth]{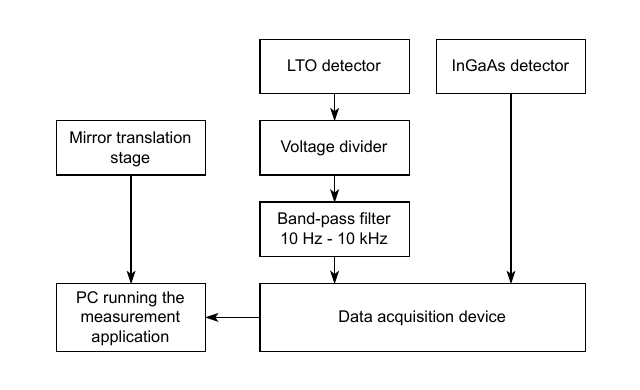} 
	\caption{Electrical and data connections.} 
	\label{fig:SetupConn}
\end{figure}

As shown in Fig. \ref{fig:SetupConn}, the measurement process is managed by a computer running custom software. Acquired IGMs and spectra can be subject to further processing and analysis. Both primary and reference detectors are connected to a~data acquisition device (DAQ) but the signal from the primary LTO detector is pre-conditioned by first dividing its voltage by 9 to fit into the dynamic range of the DAQ and then feeding it into a~Stanford Research SR560 low-noise amplifier, configured to act as a~band pass filter enforcing a 10~Hz -- 10~kHz electrical frequency window with a unitary gain. Assuming a mirror velocity equal to 1~mm/s, this frequency window translates to a~wavelength range of 400~nm~--~400~\textmu m, which provides enough margin to allow for significant speed adjustments without reconfiguration of the signal chain. The DAQ employed in the setup is Zurich Instruments MFLI operating as a~scope with the digitizer option. The noise floor of this acquisition chain was measured to be more than an order of magnitude below the noise floor of the dark LTO detector to ensure no signal degradation.  

\begin{figure}[H]
	\centering \includegraphics[width=0.85\textwidth]{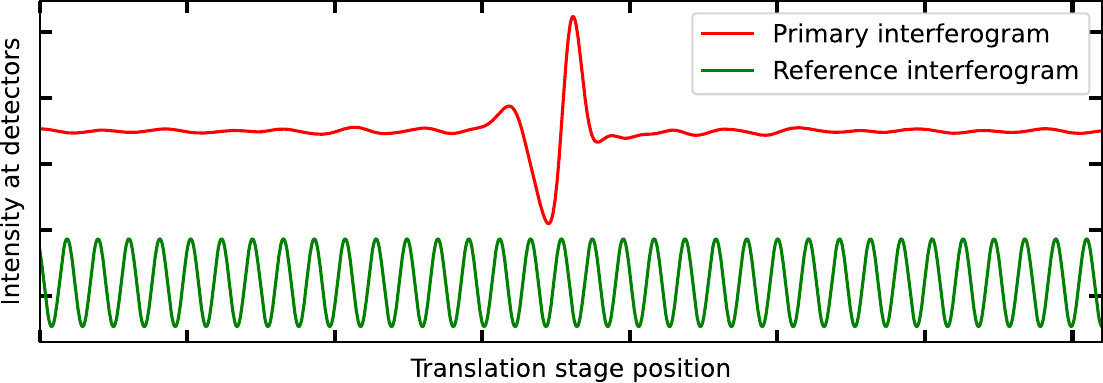} 
	\caption{Typical shapes of primary and reference interferograms as a function of the translation stage position.}
	\label{fig:interferograms}
\end{figure}

Due to imperfections in the motion of the scanning mirror, virtually all FTIRs employ an adaptive IGM sampling scheme. Although this step can theoretically be omitted with a~sufficiently precise translation stage (particularly for longer wavelengths), it is almost always present. The typical implementation relies on a~helium-neon (HeNe) laser as a~source of a~coherent beam with a stable and well-known wavelength, which is fed into the interferometer together with the primary measured beam. As the mirror is moved along the translation stage, this reference beam forms a~sinusoidal IGM along its entire path as shown in Fig.~\ref{fig:interferograms}. The reference IGM is registered by a~separate detector parallel to the primary from the thermal source. Conventionally, acquisition is triggered at characteristic waveform points (zero-crossings) to guarantee that sampling takes place only at well-defined, uniformly spaced mirror positions (optical delays) \cite{griffiths}. 

While HeNe lasers are popular sources of stable reference beams, they are too bulky to be incorporated into compact instruments and require high voltage to operate. Some commercial solutions utilize distributed feedback lasers (DFB) to address these issues~\cite{portable_spectroscopy} and in this work, we analogously perform positioning compensation using a~fibre-coupled telecommunication laser working at 1.55~\textmu m. With the traditional triggering scheme described above, the Nyquist frequency is exactly equal to it, delivering two samples per period of the reference IGM. This means clipping the lower end of spectral coverage of the instrument, so it was decided that the typical approach of triggering the acquisition at zero-crossings of the reference IGM had to be replaced with a~more precise method. Moving the Nyquist frequency far away from the normal operating range is also desirable to avoid noise folding into the registered spectrum. While it is possible to design triggering schemes generating more pulses from a~single period of the reference IGM, they would introduce jitter problems and impose a~major limitation on users. For instance, it would require the use of DAQs capable of acquiring a~single sample for every trigger event, which is not a~common feature for off-the-shelf devices. All modern scope-like DAQs can trigger an acquisition of a~preset number of samples with a~constant frequency. However, combining that with an uneven speed of the mirror translation state would result in a~significant distortion of the delay axis in the acquired IGM. 

\begin{figure}[H]
	\centering \includegraphics[width=0.9\textwidth]{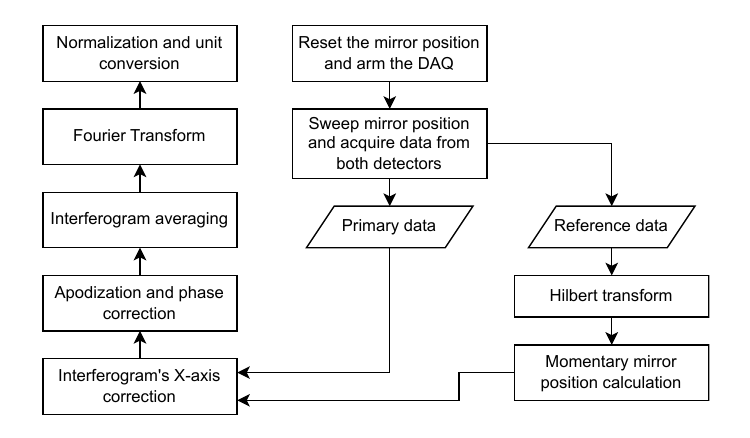} 
	\caption{Measurement process flowchart.} 
	\label{fig:MeasProcess}
\end{figure}

Instead, sample equidistance in the IGM domain is ensured here by using Hilbert Transform (HT)-based correction, in a~processing chain as visualized in Fig.~\ref{fig:MeasProcess}. HT is performed on the sinusoidal IGM $S_0^{(\mathrm{ref})}(\tau)$ of a~stable laser registered by the reference detector, where the real IGM $S_0^{(\mathrm{ref})}(\tau)$ has been converted into an analytic signal $\widetilde{S}_0^{(\mathrm{ref})}(\tau)$ via the Hilbert transform $\mathcal{H}\{\ldots\}$
\begin{equation}
\widetilde{S}_0^{(\mathrm{ref})}(\tau) = S_0^{(\mathrm{ref})}(\tau) + \mathrm{i}\mathcal{H}\{S_0^{(\mathrm{ref})}(\tau)\} \;.
\label{eq:RefIGM1}
\end{equation}
From the argument of the analytic signal (which is complex) one can directly retrieve the instantaneous optical phase lag $\varphi(\tau)=\mathrm{Arg}(\widetilde{S}_0^{(\mathrm{ref})}(\tau))=2\uppi\nu \tau$ with $\nu$ being the reference laser frequency (here 193.622~THz) and $\tau$ being the time delay between the interferometer arms. However, this poses a risk of erroneous phase unwrapping, which is required to obtain the position of discrete (non-uniform) mirror positions $L[k]$ (sampling points) via $L[k]=c\cdot\tau[k]$, where $c$ is the speed of light. Numerically, the better way is to cumulatively sum the instantaneous phase increments for the $k$-th sample $\widehat{\Delta \varphi}[k]$
\begin{equation}
\widehat{\Delta \varphi}[k] = \mathrm{Arg}\Bigg(\frac{\widetilde{S}_0^{(\mathrm{int})}[k+1]}{\widetilde{S}_0^{(\mathrm{int})}[k]}\Bigg) \,,  \, \mathrm{where} \, k\in[0,N-1] \; \mathrm{and} \; \widehat{\Delta \varphi}[0] = 0.
\label{eq:FTS_estimate1}
\end{equation}
This process avoids the problematic phase unwrapping. From the above, we can obtain the instantaneous optical phase lag via summing the instantaneous frequency in units of rad/sample
\begin{equation}
\widehat{\varphi}[k] = \sum_{i=1}^{k} \widehat{\Delta \varphi}[i] \,.
\label{eq:FTS_estimate2}
\end{equation}
By simply scaling the above results, we obtain the discrete sampling points $\tau[k]=\widehat{\varphi}[k]/(2\uppi\nu)$. This is also used to determine the maximum delay range for the $N$ samples acquired $\tau_\mathrm{max}=\widehat{\varphi}[N-1]/(2\uppi\nu)$. 

The non-uniformly-sampled thermal source IGM $S_0^{(\mathrm{int})}$ is next linearly interpolated (resampled) onto a uniform grid in an analogous fashion as in unstabilized dual-comb spectroscopy~\cite{sterczewski2019computational}. The interpolation grid yielding uniformly sampled $S_0^{(\mathrm{corr})}[k]$ is obtained from the maximum phase lag $\varphi[N-1]$. The query points ($x$-coordinates for linear interpolation) are simply
\begin{equation}
\varphi_\mathrm{q}[k] = k/(N-1) \cdot \varphi[N-1] \,.
\label{eq:interpolatioGrid}
\end{equation}
Following resampling, IGMs are apodized with the Tukey window, phase corrected using the Mertz algorithm~\cite{mertz1967auxiliary} and then averaged to a single IGM. The latter is next processed by the Fast Fourier Transform (FFT) algorithm with zero-padding to produce the final power spectrum, which after normalization and unit conversion was used for further analysis.

\subsection{Gas spectroscopy}
To enable absorption measurement of gaseous samples, a~100~mm-long glass flow cell (Pike Technologies, USA) with windows made of a~thin low-density polyethylene (PE) foil was introduced into the beam path, immediately after the collimating optics and possibly close to the interferometer input. The PE foil was selected as the window material due to a~good broadband transmission in MIR and the convenience of creating a~hermetic seal between it and the glass body of the cell. While PE absorption bands are present in the acquired spectral range, they do not strongly overlap with measured gas spectrum. 

Since both CO\textsubscript{2} and H\textsubscript{2}O are present in ambient air, the beam path needs to be purged with N\textsubscript{2}. This is problematic due to the open nature of an optical instrument assembled on a~steel breadboard. To address this challenge, the whole instrument was encapsulated in a~wrapping foil stretched onto a~3D-printed frame. N\textsubscript{2} of 99.99\% purity was pumped inside through a~diffuser until ambient gasses were removed. The chamber was considered purged after about 15 minutes of sustained high flow, after which a~spectrum acquisition was carried out without a~sample to confirm that the CO\textsubscript{2} absorption feature at 15~\textmu m can no longer be registered. Small N\textsubscript{2} flow was maintained for the entire duration of the measurement to ensure a~positive pressure difference and prevent the ambient air from entering back into the chamber.

CO\textsubscript{2} and H\textsubscript{2}O absorption measurements were performed on a~sample of human breath. Sample was diluted with N\textsubscript{2} until it could be acquired by the instrument without excessive saturation.   

\section{Results}
\subsection{Spectral dynamic range and coverage}
Two important FTS performance metrics can be extracted from a~raw spectrum of the source acquired in a purged environment: the spectral dynamic range, which is the maximum signal intensity relative to the noise floor, and spectral coverage measured between points at which the registered spectrum reaches the noise floor.

Spectra registered at higher source powers show uneven noise floors between short and long wavelength regions; therefore the spectral dynamic range is calculated relatively to the higher noise floor averaged from between 31 and 35~\textmu m.
 
\begin{figure}[t]
	\centering \includegraphics[width=1\textwidth]{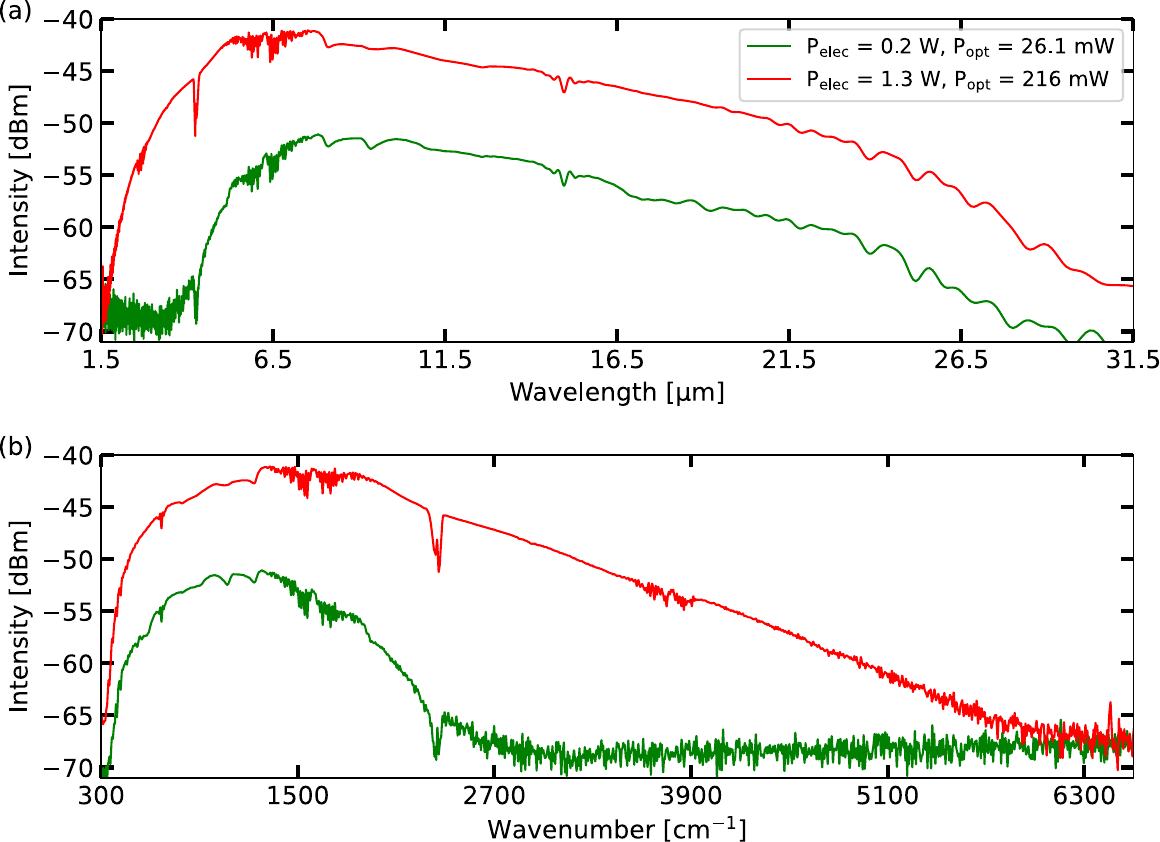} 
\caption{Spectral coverage comparison for the instrument configured with different detectors (PR~no.~1 for lower- and PR~no.~2 for higher-power setting) and the CPC source operated at two different power levels: 0.2~W input power, resulting in a~100\textdegree C surface temperature, 26.1~mW luminosity and 1.3~W input power, resulting in 360\textdegree C, 216~mW of luminosity. Panel (a) $x$-axis is scaled in wavelengths and (b) presents the same data converted to wavenumbers.
The interferograms were acquired with a 2~mm optical path difference, 2~mm/s scan speed, and 10~averages.}
	\label{fig:coverage_comp}
\end{figure}

Within the same setup, we compare two detectors using identical pyroelectric elements (LTO) with different electronics: PR~no.~1 detector with a 70~kV/W responsivity and 8~kHz bandwidth, and PR~no.~2 detector with a 5~kV/W responsivity and 50~kHz bandwidth. The highest spectral dynamic range (DR) of 23~dB was achieved with PR~no.~2 and the CPC source powered with 1.3~W. While PR~no.~1 achieved a DR of 22~dB with less than 1/3 of that power delivered to the CPC source (0.4~W), further halving that power to 0.2~W yielded a~viable spectrum with 19.5~dB dynamic range. Both measurements are compared on Fig.~\ref{fig:coverage_comp}. The spectral coverage was measured to be 1.6--31~\textmu m (9.7--190~THz) for the 1.3~W source and PR~no.~2 detector and 3.5~--~29.4~\textmu m (10.2--86~THz) for the 0.2~W with PR~no.~1. 

\subsection{Gas spectroscopy results}
\begin{figure}[t]
	\centering \includegraphics[width=0.92\textwidth]{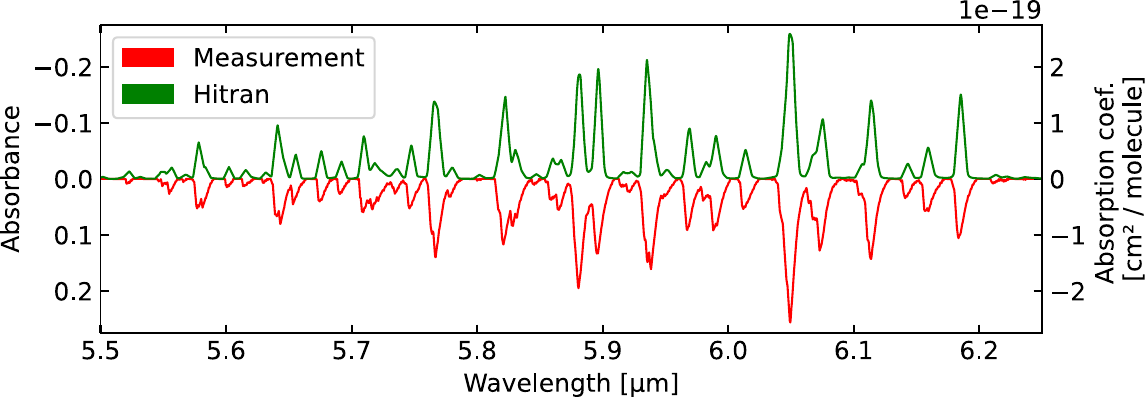} 
	\caption{H\textsubscript{2}O absorbance measured around 5.9~\textmu m compared to reference absorption coefficient from HITRAN processed to emulate a 1.4~cm\textsuperscript{-1} resolution instrument.}
	\label{fig:h2o}
\end{figure}

To demonstrate the spectroscopic capabilities of the instrument, we measured the absorption spectra of two atmospheric gas species: H\textsubscript{2}O and CO\textsubscript{2} and compared them with the HITRAN2020 database. First, H\textsubscript{2}O absorption details present between 5.5~and~6.25~\textmu m were acquired as a~spectral resolution benchmark. Absorption coefficients retrieved from the HITRAN database were processed to emulate a Michelson interferometer-based Fourier spectrometer. In other words, the simulated spectrum was apodized to mimic the effect of an instrumental line shape. Such simulated data were compared against absorbance measurements using our instrument (Fig.~\ref{fig:h2o}). All gas absorption spectra were acquired with an optical path difference of 200~mm and scan speed of 2~mm/s. Such measurements were averaged 30~times, resulting in a~total acquisition time of slightly less than 1~hour. 

\begin{figure}[t]
	\centering \includegraphics[width=0.92\textwidth]{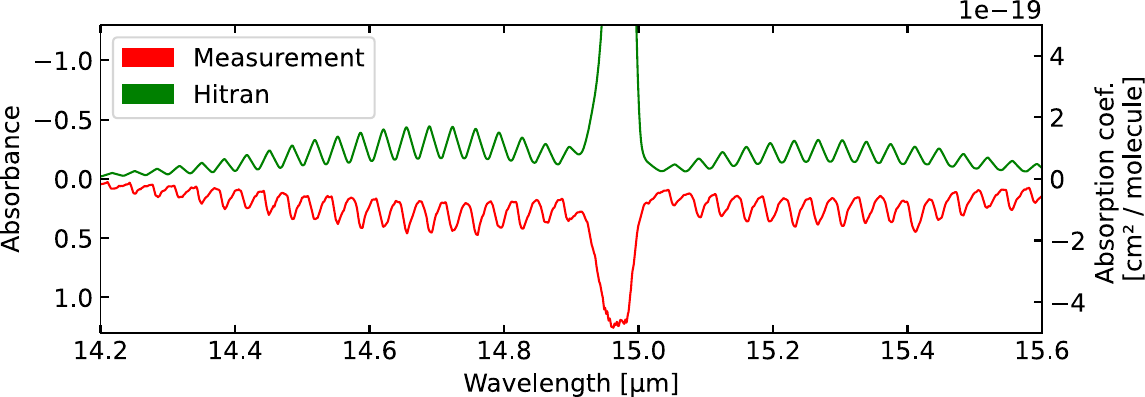} 
	\caption{CO\textsubscript{2} absorbance measured around 15~\textmu m compared to reference absorption coefficient from HITRAN processed to emulate a 0.9~cm\textsuperscript{-1} resolution instrument.}
	\label{fig:co2}
\end{figure}

\begin{figure}[t]
	\centering \includegraphics[width=0.92\textwidth]{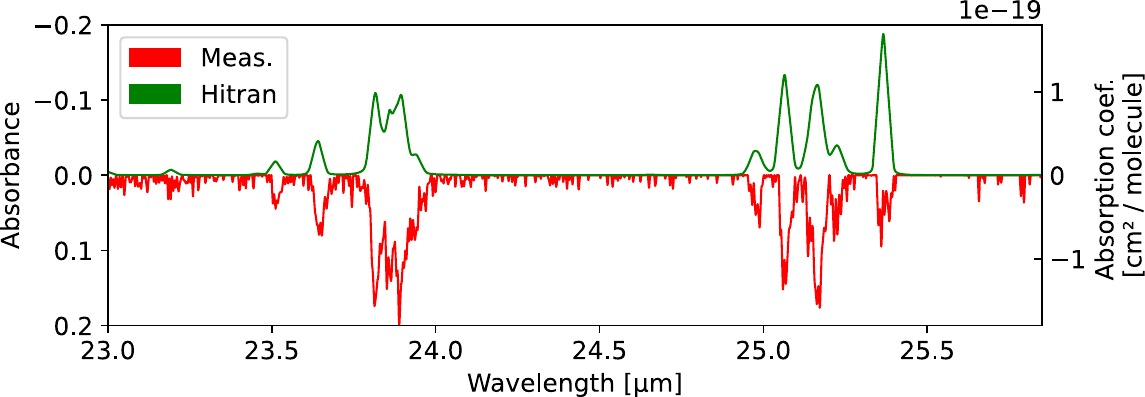} 
	\caption{H\textsubscript{2}O absorbance measured between 23~and~26~\textmu m (13--11.5~THz), compared to reference absorption coefficient from HITRAN processed to emulate a 0.5~cm\textsuperscript{-1} resolution instrument.}
	\label{fig:h2o_long}
\end{figure}

CO\textsubscript{2} absorption details present around 15~\textmu m (primarily responsible for the greenhouse effect~\cite{greenhouse}) were acquired (Fig.~\ref{fig:co2}) to demonstrate the usefulness of pyroelectric detectors in measuring wavelengths inaccessible for semiconductor devices without cryogenic cooling \cite{hamamatsu_selection}. The central peak was allowed to saturate and reach the noise floor at an absorbance of $\sim$1.25, corresponding to 5.5\% transmittance. 

Despite the lower spectral dynamic range, absorption details from H\textsubscript{2}O are still observable beyond 25~\textmu m, as shown in Fig.~\ref{fig:h2o_long}. The furthest observable peak is located at 25.35~\textmu m which corresponds to 11.8~THz. We expect an even broader coverage extending further to the far-infrared (THz) after removing the KBr window from the detector and replacing the beam splitter.

\subsection{Size, weight and power (SWaP)}
Despite the instrument's footprint of 60~$\times$~45~cm and weight of less than 20~kg, its architecture makes it well-suited for miniaturization. Power estimates from Tab.~\ref{tab:power} can be used as a~basis for battery life estimation for future portable devices. Assuming a~typical stepper motor and screw-based translation stage is used and the active-to-idle ratio is 1:10, an average power draw of such components would yield 4.6~W. This allows for >~10 hours of work from a~pack made of 4 lithium-ion 18650~cells with 3.5~Ah capacity each and 200~g weight in total. Such a long operating time is possible only thanks to the low energy consumption of the source since this is the only component that must constantly operate at full power to allow an immediate transition from standby to active state.

\begin{table}[H]
\centering
	\begin{tabular}{l c c}
		& Peak power [W] & Standby power [W] \\
        \hline
		CPC source & 1.5 & 1.5\\
		LTO detector & 0.4 & 0.0\\
        Translation stage (30~$\Omega$-coil, 12~V stepper motor) & 10.0 & 0.0\\
        Data processing module (Raspberry Pi 3B +) & 5.1 & 1.9 \\
        \hline
        Total & 17.0 & 3.4
	\end{tabular}
	\caption{Power budget estimates.}
	\label{tab:power}
\end{table}

Size and weight limitations are difficult to estimate since optical components by themselves do not contribute to them significantly. They are usually defined as design targets based on application requirements like measurement quality, resolution, and device robustness. Existing handheld FTIR would face similar limitations to solutions from this work and they can weigh less than 2.5~kg with a~footprint of 30~$\times$~20~$\times$~10~cm. Higher-resolution, laboratory devices are usually built on a~heavy steel chassis to improve their mechanical stability, easily reaching weights over 100~kg. Typical workhorse FTIRs exist in between these two extremes. Presented results can be used to achieve equivalent performance in a~smaller package due to inherent compactness and low operating temperatures of CPC sources, removal of all cooling requirements from the detector, and HT-based mirror positioning correction scheme allowing one to use smaller-footprint lasers. With the availability of optimized rotary delay line geometries~\cite{kim2023design, markmann2023frequency} or even linear piezo motors, the interferometer can in principle be realized with a~centimeter-scale footprint.

\section{Conclusions}
The demonstrated spectral coverage of 1.6~--~31~\textmu m (322~--~6250~cm\textsuperscript{-1}) with 23~dB of dynamic spectral range is competitive to commercial solutions and enough to satisfy requirements in many areas including biological tissue analysis~\cite{tissues_FTIR} or food authenticity and adulteration analysis~\cite{food_FTIR}. Our results were achieved with a~single optical configuration (without changing the detector or beam splitter) and at room temperature with an uncooled detector and source. 

With the large emission surface area (5 $\times$ 5 mm\textsuperscript{2}) of the CPC source comes an expectation of a~degraded resolution. This is because the further a~source is from an ideal point, the worse collimation parabolic mirrors can produce. The typical way of improving it is to include collimating optics with a J-Stop to enable trading spectral dynamic range for improved resolution. In this work, we show that 1.5~cm\textsuperscript{-1} can instead be achieved with a much simpler setup with only a~single OAP mirror performing the collimation. This result is more than satisfying for acquiring spectra of condensed-phase samples since widths of their absorption bands are rarely narrower than 2.5~cm\textsuperscript{-1}~\cite{griffiths}. 

The estimated resolution of the instrument differs for the supported spectral regions: 1.4~cm\textsuperscript{-1} for 6~\textmu m, 0.9~cm\textsuperscript{-1} for 15~\textmu m and 0.5~cm\textsuperscript{-1} for 25~\textmu m. Improvement in resolution for higher wavelengths can generally be expected due to imperfect collimating optics and mechanically unstable translation stage which affect short wavelengths more than long ones. Since the presented instrument uses a traditional screw-stepper motor drive and flat mirrors, its prone to interferogram modulation errors resulting in lower SNR and deformations of the spectrum \cite{fts_instrumentation}. 

The mechanically-unstable translation stage is the primary limitation of the scan velocity. With a better mechanical setup, PR~no.~2 detector allows for \textgreater~10$\times$ faster acquisition times without significant degradation in performance, since the shortest acquired wavelength of 1.6~\textmu m corresponds to 2.5~kHz, being far from the optimal utilization of its full 50~kHz bandwidth.

Mirror positioning error correction based on simultaneous acquisition of the reference laser interferogram and then performing HT on it to retrieve momentary position information proved to be an effective solution in the investigated wavelengths range. It remains unclear where the limit of this technique lies with regard to the shortest wavelength from a broadband source. This work proves that this scheme can enable further miniaturization of instruments thanks to replacing traditional HeNe lasers with telecommunication semiconductor ones operating at 1.55~\textmu m, which can offer good noise performance and thermal stabilization in a thumb-sized package. While these devices struggle to match HeNe emission stability, they offer about 1000$\times$~higher power efficiency, do not require high-voltage power supplies and occupy a small volume. Thanks to the wide adoption by the telecom industry, these devices can benefit from effects of scale therefore being rugged, highly-standardized and cost effective. The major downside of this approach is a~significant increase in demand for computing power, since performing the HT-based correction is a more computationally demanding task than just the FFT. The relatively recent emergence of high-performance, power efficient and affordable single-board computers like Raspberry~Pi, BeagleBone, etc. which can easily handle tasks as demanding as real-time image analysis, suggests that computational resources should not be a major obstacle to a wider adoption of the discussed technique~\cite{rpi_vs_stm32, review_of_pi_applications}.

Thin-membrane LTO detectors, with their detectivity approaching DLATGS, high modulation bandwidth, excellent linearity and thermal stability, emerge as a~technology-of-choice for future generations of compact, affordable FTIR spectrometers. This work proves their parameters are already sufficient for practical measurements. The low thermal coefficient of LTO makes it a~prime candidate for manufacturing low-thermal mass, thin-membrane detectors as the heating effect from the detected radiation can become very significant. Despite the low thickness of their active area, such detectors are known to survive shocks of 5000~g~\cite{stokowski_ion-beam_1976}.  

Close to an ideal black body, the broadband emissivity of CPC sources allows them to not only achieve significant power savings, but also operate at lower temperatures than typical globars, while maintaining enough optical output power to perform practical spectroscopy. As shown in Fig.~\ref{fig:coverage_comp}, even with electrical power as low as 0.2~W, corresponding to a~source temperature of about 100~C\textdegree and 26.1~mW of optical power, the setup was still able to render quality broadband MIR spectra with almost 20~dB of dynamic range. This property is particularly valuable for applications in biology, organic chemistry and medicine where too much heat delivered from the source to the sample can disturb or damage it. The maximum operating temperature of the tested CPC source is about 1000\textdegree C, significantly lower than typical working points of globars between 1000~and~1600\textdegree C. This is expected to result in longer lifespans, shorter startup times and a~complete removal of the need for water cooling. 

The evaluated prototype of KBr-Ge beam splitter with an integrated dispersion compensation plate and anti-reflective coating allows one to design more compact spectrometers and helps to mitigate etaloning, the lack of which is evident in Fig.~\ref{fig:coverage_comp}. No attempt was made to numerically remove etalons from these spectra. They could still be observed in higher-resolution scans of gas samples, especially at longer wavelengths, but their depth was minimal at \textless1~dB at~25~\textmu m and did not result in periodic regions of excess noise appearing in acquired gas spectra. Such beam splitter construction also decreases the total number of elements in the beam path, thus making it easier to align and less prone to later drift.

\section{Outlook}
We expect the presented FTIR setup to initiate the construction of transportable, possibly handheld, instruments capable of covering the entire MIR spectral range as well as parts of NIR and THz. With all cooling requirements removed and Watt-level consumption of NTSs, the translation stage and the main computer become primary energy consumers. This should allow for a whole work day operation with a~lightweight battery pack. Appearance of cost-efficient detectors tends to be followed by the introduction of affordable spectroscopy solution for their respective wavelength coverage~\cite{portable_spectroscopy} and authors believe thin-membrane LTO detectors to be at a sufficient development level to enable this transition for FTIR, especially when combined with low-power NTS. Compact thermal sources are expected to evolve further in the direction of greater power efficiency thanks to metamaterial-based solutions, like selective thermal emitters like those described in \cite{metamaterial_black_body}.

These results synergize well with current trends of employing multivariate analysis and machine learning algorithms for extracting information from complex mixtures and materials, since their performance increases together with the spectral bandwidth~\cite{food_FTIR, FTIR_coffee_spices}. Such intelligent devices would also benefit from the availability of a compact, more affordable and rugged platform which would be much easier to manufacture and deploy in different environments, initially to gather large datasets required for calibration of these algorithms and later to make them more attractive to customers. It is also  important to note that such portable, intelligent instruments should not be seen as competition to professional, laboratory-grade equipments, but rather as means to deliver spectroscopy to new fields \cite{portable_spectroscopy}. 

In the concluding paragraph, we would like to note that the finite delay range is not a fundamental resolution limit of this instrument ($\sim$GHz). With optical frequency combs, it is possible to surpass it a thousand times to obtain kHz optical resolutions~\cite{maslowski2016surpassing} using a specialized IGM sampling routine. Using chip-based frequency combs in the MIR~\cite{sterczewski2024sub}, MHz resolution at millimeters of optical path difference should be easily accessible with this instrument with a large wavelength agility.

\vspace{0.5cm}

\noindent\textbf{Funding} European Union (ERC Starting Grant, TeraERC, 101117433).
\vspace{0.3cm}

\noindent\textbf{Acknowledgements.} The authors acknowledge funding from the European Union (ERC Starting Grant, TeraERC, 101117433). Views and opinions expressed are however those
of the authors only and do not necessarily reflect those of the European Union or the European Research Council Executive
Agency. Neither the European Union nor the granting authority
can be held responsible for them. The authors thank Mihai Suster and Aleksandra Szymańska from Faculty of Physics at University of Warsaw for their help in producing earlier versions of beam splitters. This work is supported by the use of National Laboratory for Photonics and Quantum Technologies (NPLQT) infrastructure, which is financed by the European Funds under the Smart Growth Operational Programme. 
\vspace{0.3cm}

\noindent\textbf{Disclosures.} J. K. is an employee of LASER COMPONENTS Germany GmbH offering the pyroelectric detectors and beam splitters used in this study in partnership with WiredSense GmbH owned by M. B. and T. G.. M. S. is an employee of Infrasolid GmbH offering the broadband thermal source. No other conflict of interest is present.
\vspace{0.3cm}

\noindent\textbf{Data availability.} Data underlying the results presented in this paper are available in Ref.~\cite{Mnich2024}.

\bibliography{literature}

\end{document}